\documentclass[sigconf]{acmart}

\usepackage{epsfig}
\usepackage{booktabs} 
\usepackage{multirow}
\usepackage{xspace}
\usepackage{enumitem}
\usepackage{array}
\usepackage{makecell}
\usepackage{graphicx}

\def\showcomments{}
\ifdef{\showcomments}
{
\newcommand{\ye}[1]{\textcolor{blue}{$\ll$\textsf{#1 --YE}$\gg$}}
\newcommand{\hr}[1]{\textcolor{cyan}{$\ll$\textsf{#1 --HR}$\gg$}}
\newcommand{\ol}[1]{\textcolor{red}{$\ll$\textsf{#1 --OL}$\gg$}}
\newcommand{\an}[1]{\textcolor{brown}{$\ll$\textsf{#1 --AN}$\gg$}}
}
{
\newcommand{\ye}[1]{}
\newcommand{\hr}[1]{}
\newcommand{\ol}[1]{}
\newcommand{\an}[1]{}
}

\newcommand{\ecom}{{e-commerce}\xspace}
\newcommand{\Ecom}{{E-Commerce}\xspace}
\newcommand{\ebay}{{eBay}\xspace}
\newcommand{\model}{{PreSizE}\xspace}

\newcommand{\squeezeup}{\vspace{-2.5mm}}
\newcommand{\squeezeupabit}{\vspace{-1.5mm}}

\copyrightyear{2021}
\acmYear{2021}
\setcopyright{acmcopyright}
\acmConference[SIGIR '21]{Proceedings of the 44th
International ACM SIGIR Conference on Research and Development in Information
Retrieval}{July 11--15, 2021}{Virtual Event, Canada}
\acmBooktitle{Proceedings of the 44th International ACM SIGIR Conference on
Research and Development in Information Retrieval (SIGIR '21), July 11--15, 2021,
Virtual Event, Canada}
\acmPrice{15.00}
\acmDOI{10.1145/3404835.3462844}
\acmISBN{978-1-4503-8037-9/21/07}

\begin{document}
\fancyhead{}

\title{\model: Predicting Size in \Ecom using Transformers}

\author{Yotam Eshel, Or Levi, Haggai Roitman, Alexander Nus
}
 \affiliation{
   \institution{eBay Research}
   \city{Netanya}
   \country{Israel}
}
\email{\string{yeshel, olevi, hroitman, alnus\string}@ebay.com}

\renewcommand{\authors}{Yotam Eshel, Or Levy, Haggai Roitman, Alexander Nus}

\begin{abstract}

Recent advances in the \ecom fashion industry have led to an exploration of novel ways to enhance buyer experience via improved personalization. Predicting a proper size for an item to recommend is an important personalization challenge, and is being studied in this work. 
Earlier works in this field either focused on modeling explicit buyer fitment feedback or modeling of only a single aspect of the problem (e.g., specific category, brand, etc.). More recent works proposed richer models, either content-based or sequence-based, better accounting for content-based aspects of the problem or better modeling the buyer's online journey. However, both these approaches fail in certain scenarios: either when encountering unseen items (sequence-based models) or when encountering new users (content-based models).

To address the aforementioned gaps, we propose \model\ -- a novel deep learning framework which utilizes Transformers for accurate size prediction. \model models the effect of both content-based attributes, such as brand and category, and the buyer's purchase history on her size preferences. Using an extensive set of experiments on a large-scale \ecom dataset, we demonstrate that \model is capable of achieving superior prediction performance compared to previous state-of-the-art baselines. 
By encoding item attributes, \model better handles cold-start cases with unseen items, and cases where buyers have little past purchase data. As a proof of concept, we demonstrate that size predictions made by \model can be effectively integrated into an existing production recommender system, yielding very effective features and significantly improving recommendations.
\end{abstract}



\maketitle
\section{Introduction}
\label{sec:intro}

The growth of the \ecom fashion industry has driven forward a large body of research on new personalization problems. Among such problems is \emph{size prediction}, which is the focus of our work. 

Correctly predicting the size preferences of e-commerce buyers when recommending items can improve buying experience and result in less item returns. 
However, predicting the right size in \ecom is not a trivial task for multiple reasons. First, the notion of size by itself is ambiguous due to the fact that there are different sizing schemes (e.g., \texttt{EU}, \texttt{UK}, \texttt{US}, etc.), different scales (e.g., numerical, (\texttt{S},$\ldots$,\texttt{XL}), (\texttt{A},$\ldots$,\texttt{DD}), etc.) and different usages of the term size (`\texttt{shoe size}', `\texttt{bottoms size}', `\texttt{hosiery size}', `\texttt{cup size}', etc.).
In addition, there are no clear size conversion charts, and size may vary between different brands (e.g., UK size 10 can be converted to both US size 6 and US size 8, depending on the brand).
Moreover, buyers' own size preferences may change over time. Figure~\ref{motivation} illustrates such a real example taken from our data. In this example, the buyer's size preference for kids clothing has changed over a period of six months. As this example demonstrates, size preferences can be quite dynamic and even dramatically change over a short course of time. 
Lastly, a single account may be used by multiple buyers, having a multitude of size preferences, making the prediction task even harder.

\begin{figure}
\small
\begin{tabular}{lll}
\begin{tabular}[c]{@{}l@{}}\textbf{Date}\end{tabular} & \textbf{Category} & \begin{tabular}[c]{@{}l@{}}\textbf{Size}\end{tabular} \\
\toprule
 2019-12 & Baby/Toddler:Outfits \& Sets & 3-6 months \\
2020-05 & Baby/Toddler:Outfits \& Sets & 12-18 months \\
2020-05 & Baby/Toddler:Outfits \& Sets & 3T \\
2020-05 & Baby/Toddler:Tops \& T-Shirts & 24 months \\
2020-05 & Baby/Toddler:Outfits \& Sets & 24 months
\end{tabular}\caption{Example of a dynamically changing size preference over a course of six months based on a given buyer's purchase history in toddler categories. }\label{motivation}
\end{figure}


The size prediction task has been addressed by several previous works in recent years and various solutions have been suggested. Earlier works have either relied on explicit buyers' fitment feedback or have trained specific prediction models that cover only a single aspect of the problem (e.g., a single brand, category, buyer segment, etc.).
Yet, in a real-world \ecom setting, buyers' fitment feedback is usually scarce; therefore, we utilize implicit feedback based on buyers' shopping journeys instead. Additionally, training a model for every brand or category does not scale well, considering the variety of merchandise that may be offered online. More recent work addressed a more realistic setup and proposed methods to model multiple aspects of the problem. However, these works still suffer from cold-start issues, lacking the ability to handle either new buyers or novel items. Finally, almost all previous works do not explicitly model the temporality of a buyer's shopping journey. Such a journey is commonly characterized by a sequence of items that were purchased by the buyer prior to the next item purchase time whose size we wish to predict. 


Following previous work, we formulate the size prediction task as a multi-class classification problem over all size values which appear in the dataset. We propose \textbf{\model}\footnote{\model\ stands for ``Predicting Size in \Ecom''.} -- a novel size prediction framework that utilizes Transformers~\cite{vaswani2017attention} in two main ways. First, we use a Transformer to capture the relationship between various item attributes (e.g., brand, category, etc.) and its purchased size. This Transformer basically co-embeds each item's size property with its associated attributes. This in turn, allows to address cold-start cases, where learned size and attribute embeddings can be shared among items (and buyers) for better handling novel (unseen) items.  
Second, we utilize a second Transformer layer which captures the relationship between previous item purchases in a given buyer's history and the size preference of the next item being bought. 

Most previous works have utilized relatively small scale public datasets that include explicit buyer's fitment feedback (e.g., Small, Fit, Large). Others have utilized in-house datasets while focusing only on specific item subsets (e.g., specific category or buyer segment). Instead, in this work, we use a proprietary large-scale dataset from the \ebay\ \ecom website, which is based on implicit feedback obtained from buyer purchase histories. Our dataset is unique, having both strong buyer-item usage sparsity and loose limitations on the sellers' input. Overall, we evaluate \model over items with rich attributes, spanning over multitude of departments (e.g., `\texttt{Men's}', `\texttt{Women's}', etc.), item types (e.g., `\texttt{Tops}', `\texttt{Bottoms}', etc.) and buyer account types (e.g., `\texttt{Single Gender}', `\texttt{Mixed Age}', etc.). 



We evaluate \model against previous state-of-the-art baselines and demonstrate its superior size prediction quality. Moreover, items in our dataset may have several size options and the size relevant to a specific buyer may not always be available. Using another large-scale \ebay dataset, we further demonstrate that, \model size predictions can be utilized as features for enhanced personalization, helping to capture the likelihood that an item will have available inventory with the relevant size for a given buyer. By doing so, we improve an existing production item recommendation service by a significant margin.  

Overall, our contributions can be summarized as follows:
\begin{itemize}
    \item We present \model -- a novel approach to predict buyers' size preferences in \ecom based on 
    item purchase histories.
    \item Using a large-scale dataset from \ebay website, we demonstrate the merits of \model,  significantly outperforming all previous approaches which tackled the same task.
    \item We show the impact of size prediction on a real item recommendation service deployed on the \ebay website.
    
\end{itemize}

\section{Related Work}
The size prediction task is a relatively new task and has been previously studied by several related works~\cite{AttentionRecsys2020, dogani2019learning, lasserre2020meta, Misra2018, Guigoures2018, Sheikh2019a, singh2018footwear, Sembium2017, Sembium2018}. We next briefly review related work, further emphasizing the main differences from our work. 

A first line of related works~\cite{Guigoures2018,Misra2018, Sembium2017,Sembium2018} focused on predicting size fitment from explicit buyer's fit feedback (e.g., Small, Fit, Large). Among these works, Sembium et al.~\cite{Sembium2017} have cast the size prediction task
as an ordinal regression problem, where differences between true buyer and item sizes were fed into a linear model.  The same authors~\cite{Sembium2018} have extended their solution using a Bayesian regression model with ordinal categories. Guigour\`{e}s et al.~\cite{Guigoures2018} have proposed a hierarchical Bayesian model that learns the joint probability of a buyer purchasing a given item size and its fitment. Misra et al.~\cite{Misra2018} have utilized a combination of ordinal regression and metric learning for better handling of class imbalance. 

A second line of related works, sharing a 
similar problem setting, 
are those that have utilized attributes of buyers and their purchased items (e.g., category, brand, chest size, length, etc.) as implicit feedback for size prediction~\cite{Abdulla2017SizeRS, dogani2019learning,Sheikh2019a,lasserre2020meta,singh2018footwear}. Such works are motivated by the fact that, explicit customer fit feedback is usually noisy (e.g., based on customer free text feedback on returned items) and sparse.

Among these works, 
Dogani et al.~\cite{dogani2019learning} have proposed the \emph{Product Size Embedding} (PSE) neural collaborative filtering model. Within this model, item embeddings were learned for each possible size. Using an asymmetric modeling approach, buyers were then represented by items in their purchase history.  
Abdulla and Borar~\cite{Abdulla2017SizeRS} trained a classifier for fitment prediction using a combination of observable and latent buyer and item features. Latent features were obtained using a skip-gram model learned over buyer purchase histories. The same model was extended in~\cite{singh2018footwear} for footwear size recommendation and enhanced with a probabilistic graphical model
that considered brand similarities. 

A notable limitation of all the aforementioned works, 
is the requirement to train a model for a specific category, brand or size, which does not scale well in real \ecom settings, where a high variety of merchandise is common. 
Two recent works have further tried to overcome this limitation~\cite{Sheikh2019a,lasserre2020meta}. Sheikh et al.~\cite{Sheikh2019a} have proposed \emph{SFNet} - a deep-learning based content-collaborative model for personalized size and fit recommendation. Similar to our work, the size prediction task was modeled as a multi-class classification problem trained over historical buyer-item interactions. Buyer and item latent embeddings were derived using a combination of embedding content features and applying feed-forward layers with skip connections. Lasserre et al.~\cite{lasserre2020meta} have utilized a meta-learning approach, where buyers were represented by items in their purchase history and their attributes. Using embedded linear regression, both items and their purchased sizes were mapped into a latent space where they share a strong linear dependence. Yet, both~\cite{lasserre2020meta,Sheikh2019a} and previously mentioned works so far do not consider the sequence of item purchase events in a given buyer's history. As was illustrate in Figure~\ref{motivation}, such modeling can allow to capture additional sequential patterns that better represent the buyer's size preferences. 


In recent years, Transformers~\cite{vaswani2017attention} have been highly adopted for their success in prediction over sequential tasks.  
 Transformers have been initially applied with a great success in NLP tasks (e.g., BERT~\cite{devlin2018bert}, GPT~\cite{radford2019language} and XLNet~\cite{yang2019xlnet}). In the recommendation systems domain, Transformers have been primarily applied in sequence-based recommendation tasks~\cite{chen2019behavior,Fei2019,Fischer2020,Wu2020}. Given a sequence of item identities in the buyer's history, Transformers were utilized for predicting the identity of the next buyer-item interaction. Yet, the goal of such prediction tasks is eminently different from that of the size prediction task. 

To accommodate the sequential dependency in size prediction, in this work, we also utilize Transformers. Similar to our approach, Hajjar and Zhao~\cite{AttentionRecsys2020} have recently utilized Transformers to encode the sequence of buyer's purchases based on purchased item identities and their sizes. Yet no additional buyer or item attributes are considered in~\cite{AttentionRecsys2020}, which limits generalization into never-seen-before items (cold-start). In contrast, we use additional Transformer layers to generate item embeddings from their attributes without relying on item-ids. Using such additional Transformer layers allows to better capture the effect of item aspects (e.g., title, category, brand, etc.) on purchased item sizes and generalize into unseen items.

Finally, we further demonstrate the utilization of size prediction as an important feature for enhanced personalization, by training 
a recommendation model with size predictions outputted by our model. To the best of our knowledge, no prior work has explored this research direction. 


\section{Size Prediction Framework}
\label{sec:models}

 In this section we describe the details of our \model size prediction framework. We first formulate the problem as a multi-class classification problem. We then describe the \model\ model architecture and implementation details. We conclude this section with a short discussion of how \model's size predictions can be utilized as features within a downstream item recommendation task. 

\subsection{Problem Formulation}
Let $U$ denote a set of buyers and let $I$ denote a set of items. 
For a given buyer $u\in{U}$ and her previously purchased items $H_u$ (hereinafter referred to as the buyer's purchase history), we formulate the size prediction problem as estimating $P(s|i,H_u)$, i.e., \emph{the probability that the next item $i\in{I}$ that will be purchased by buyer $u$ will be of size $s$}. 


While a size variable $s$ can be thought of as a continuous or ordered variable, in practice, sellers may use a mixture of overlapping, incompatible and discrete measurement systems which are not easily interpreted. As an example, lets consider numeric US and UK clothing sizes. Both size systems are ranging from 0 to low 20's. However, some conversion tables will show that a UK size 10 should be converted to a US size 6, whiles others to 8. Sellers often do not specify the size system they use in a structured way, and even when they do, universal conversion tables themselves often disagree. In addition, enforcing even a partial ordering on sizes can be labour intensive and error prone. 

To overcome such sizing complications, 
we treat size $s$ as a categorical variable. 
We collect size measurements as strings, assigning a unique id to each unique string. 
By using this approach, our size prediction problem becomes a multi-class classification problem. 

We solve the size prediction problem by defining a model $f$, which is implemented in this work as a deep neural-network. Our goal, is to train $f$, such that the difference between the model's size prediction and the observed size is minimized. 

Formally, during train time, for a given item $i\in H_u$, let $H_{u[\prec{i}]}$ denote the sequence of items purchased by buyer $u$ prior to item $i$'s purchase time. Let $\hat{y}_{s_i}=f(H_{u[\prec{i}]},i)$ and $y_{s_i}$ further denote item $i$'s estimated and ground truth labels respectively. 

We aim at minimizing the overall miss-classification loss:

\begin{equation}
    Loss(U,I)=\sum_{u \in U}\sum_{i \in H_u} L(\hat{y}_{s_i}, y_{s_i}),
\end{equation}

where $L(\hat{y}_{s_i}, y_{s_i})$ is calculated as the cross-entropy loss: 


\begin{equation}
L(\hat{y}_{s_i}, y_{s_i}) = -\sum_{j}y_{s_i}[j]\log(\hat{y}_{s_i}[j])
\end{equation}

\subsection{Model Architecture Overview}
\label{sec:arch}

\begin{figure*}
  \centering
  \includegraphics[scale=0.9]
  {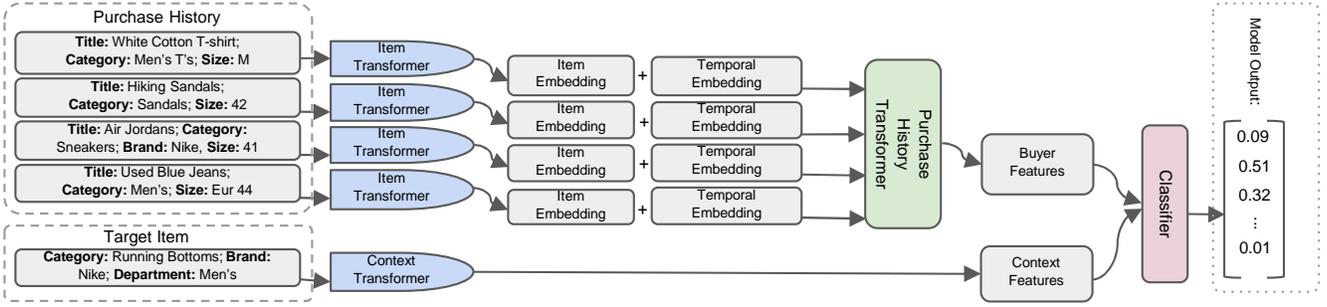}
  \caption{Overview of \model architecture. (1) Item Transformer is used to obtain a representation of each item in the buyer's purchase history. (2) Second Transformer is used to obtain a representation of the buyer's entire purchase history. (3) Context Transformer (weight shared with Item Transformer) is used to represent the target item. (4) Classifier network is used to predict the most likely purchased size of the target item given buyer's purchase history.}
  \label{fig:Model}
\end{figure*}

The architecture of the \model\ deep neural-network is illustrated in Figure~\ref{fig:Model}. The network is composed of four main steps. In the first step, we embed each item in the buyer's purchase history and obtain its dense feature representation (embedding). Then, using the item embeddings along with their purchase times, we obtain a dense feature representation of the whole buyer’s purchase history. In parallel, we produce a context embedding of the target item for size prediction. Finally, both the buyer's purchase history and context embeddings are fed into a classifier network which outputs the size (class) probability estimates. 

During training, the buyer's (purchase history) embedding sub-network learns to extract features describing the general size preferences of a particular buyer. On the other hand, the context embedding sub-network learns to extract features that allow the model to adjust such general size preferences to a particular item of interest. In this work, we obtain the various embeddings using Transformers~\cite{vaswani2017attention}, whose usage details will be detailed in Section~\ref{model implementation}.  

A notable property of our architecture is that, while our model is trained end-to-end, the item embedding computation is independent for each item. Since most of the computation is done during item embedding, we can pre-compute such embeddings in an offline process. Then, during inference, we fetch relevant item embeddings and only compute the buyer and context embeddings on-the-fly. This is particularly attractive in a production environment, since inference can be done efficiently online without compromising on up-to-date buyer purchase histories.  

\subsection{Handling Data Sparsity}\label{sec:sparsity}

A distinctive characteristic of our setting is high sparsity of buyer purchase histories and of the purchased items themselves (as will be further discussed in Section~\ref{dataset_section}). This fact entails that, during test time, we are likely to encounter many little-seen buyers and never-seen-before items. Such a setting has challenged previously proposed models~\cite{AttentionRecsys2020} that rely on embedding item ids and buyer ids, and require enough training examples per item and buyer to learn useful latent features. To address this challenge, we encourage our model to generalize based on explicit content features of items and buyers, rather than learning latent features directly.  We, therefore, represent each buyer $u\in{U}$ by the sequence of purchased items in $H_u=((i_1,d_1),\ldots,(i_j,d_j),\ldots)$, ordered by their purchase times $d_j$. Each item $i\in{H_u}$ is further represented by its set of attribute name-value pairs: $i=\{(n_{i1},v_{i1}),\ldots,(n_{il},v_{il})\}$. We assume that the set of all possible attribute names is fixed, while attribute values may consist of arbitrary string lengths. In practice, our attributes include the item’s \texttt{title}, \texttt{category}, \texttt{size measurements} and 11 other features such as \texttt{brand}, \texttt{country of manufacture} and \texttt{department}.  More details on the exact features will be discussed in Section~\ref{dataset_section}. 

\subsection{Model Implementation}\label{model implementation}
We next describe in detail the implementation of the various modules in the \model neural-network. 

\subsubsection{Item Embedding}
\label{sec:item_embd}
\squeezeup
\begin{figure}[b!]
  \centering
  \includegraphics[scale=0.65]{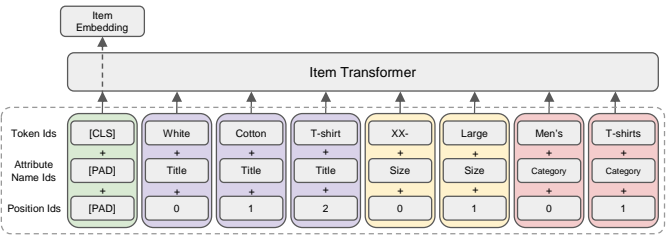}
  \caption{Generating content-based item embeddings. An item is represented as a set of attribute name-value pairs. Each attribute value is separately tokenized. Token sequences are then concatenated and fed to a Transformer along with positional and attribute name id embeddings.}
  \label{fig:Item}
\end{figure}

We first produce a dense feature representation of each item in the buyer’s purchase history.
A schema of this step is depicted in Figure~\ref{fig:Item}. Since we represent an item by its set of name-value attributes, and the length of attribute values may vary, we first tokenize the attibutes. For a given attribute value $v$, let $t_{1},\ldots,t_{|v|}$ denote its sequence of token ids. We further extend each token id with its relative position $j\in[1,\ldots,|v|]$ and the associated attribute-name id $n$ to obtain a series of triplets: $(t_{1},1, n)\ldots(t_{|v|},|v|, n)$.
We then concatenate all triplet sequences to a single sequence describing the entire item. We use an embedding layer to embed each component of a given triplet $(t_{j},j, n)$ into corresponding dense vectors $h_{t_{j}}$, $h_{j}$ and $h_{n}$. The triplet's embedding is then obtained by summing up its sub-components' embeddings (i.e., $h_{t_{j}}+h_{j}+h_{n}$). 
Following standard practice, we prepend a ‘\texttt{[CLS]}’ token embedding to this sequence and feed it into a stacked Transformer layer. We consider the Transformer’s output corresponding to the ‘\texttt{[CLS]}’ token as the resulting item's embedding. 

A Transformer is a widely used state-of-the-art attention-based model for processing textual input. We omit a detailed description of the Transformer model for brevity. For a detailed description, the reader is kindly referred to~\cite{vaswani2017attention}. However, a noteworthy property of the attention model is that it is invariant to input order and requires positional embeddings to know the order of tokens in a sequence. We use this to our advantage by providing positional embedding relative to each attribute-value sequence only, supplemented by attribute (name) id embeddings. This allows our model to know which attribute each token belongs to and the ordering of tokens within each attribute value, but does not imply any artificial ordering of the attributes. 

\subsubsection{Buyer Embedding}
\label{sec:buyer_embd}

In the second step of our model we obtain a representation of the buyer's purchase history $H_u$. To this end, we introduce temporal-embeddings so our model is made aware of the ordering and time-based relevancy of past purchases. To obtain the temporal embeddings, we devise a simple method that avoids sparsity, yet allows our model to identify which purchases are more recent and which are further in the past. For each item $i\in{H_u}$, we first compute $d'_i=d_{ref}-d_i$: the days that elapsed between item $i$'s purchase time $d_i$ and the current target purchase time $d_{ref}$\footnote{We note that, during training, the purchase time of the last item in $H_u$ is used as the reference time $d_{ref}$.}. We then compute $floor(log(d'_i))$ by rounding down the log of $d'_i$ and consider this as its temporal id. This formulation allows the model to consider the scale of time elapsed since item $i$'s purchase time, and therefore its time-based relevancy, while eliminating sparsity. We use an embedding layer to turn temporal ids into dense vectors.

Once we obtain the temporal embedding for each item in $H_u$, we combine it with the item's embedding from the previous step by summing both embeddings together. To obtain the representation of the whole purchase history, we feed the sequence of the purchased item embeddings into a second stacked Transformer layer. We further prepend a ‘\texttt{[CLS]}’ token to our sequence and consider the Transformer’s output corresponding to the ‘\texttt{[CLS]}’ token as the representation (embedding) of the entire buyer's purchase history.

\subsubsection{Context Embedding}
\label{sec:context_embd}

In the third step, we produce the context (features) representation for the target item. Context features allow our model to adjust the general size preferences of a given buyer into a size measurement that fits the particular target item. In practice, this step is identical to producing item embedding, with the exception that all the actual size measurements and the title of the target item are masked out. This masking is required to prevent label leakage during training, when our model is optimized to estimate the actual purchased size of the target item. The embedding and Transformer layers used for this step are weight-shared with those used in the item embedding step.

\subsubsection{Classifier}
\label{sec:clasifier}

In the final step, we obtain the size predictions. Here, we first concatenate the buyer's purchase history and context embeddings. We then apply a 3-layer feed-forward network which transforms the concatenated embeddings into class probability estimates. We use Gaussian Error Linear Unit~\cite{hendrycks2016gaussian} (GELU) non-linearities after the first two hidden-layers and a softmax layer to compute the final class probabilities after the third layer.

\subsection{Utilizing Predicted Sizes as Features}
\label{sec:features}
We conclude this section by making an observation that, apart from only predicting the right size of a given target item (for a given buyer), \model's predictions can be further utilized as personalization features for downstream  tasks, such as item recommendation. 
In a recommendation task, we are given a buyer $u$, and an item $i$, where $i$ can have multiple size options the buyer might purchase. 
We are interested in using \model to generate features describing the likelihood of the available sizes of $i$ will fit $u$'s size preferences. These features can be then used by a recommender system to improve its recommendations. We expect such features to be extremely helpful in cases where item size inventory is only partial. 

To generate such features, we first view the output scores of \model's final softmax layer as a probability distribution over the sizes a buyer is likely to buy of a given item. We then propose three size-prediction based features, namely: \emph{total score}, \emph{best score} and \emph{best rank}. The \emph{total score} feature represents the summation of \model predictions over the available size inventory, capturing the probability that at least one size will match the buyer's preference. The \emph{best score} feature is calculated by computing the size option with maximum \model predicted probability, assuming the buyer would prefer the best option. Finally, the \emph{best rank} feature provides a smoothing over the likelihoods by replacing the score of the most probable available size with it's rank among all possible sizes. As we demonstrate in our experiments (Section~\ref{recommend with size}), utilizing such additional size features within a downstream learning-to-rank setting, improves item recommendation.

\section{Experiments}
To demonstrate the usefulness of our \model framework, we conduct a wide array of experiments on data collected from the \ebay\ \ecom website. We start by describing our experimental setup. We then present our results for the main task of size prediction. We analyze our results across several important dimensions and perform an ablation study. We conclude this section with a proof of concept, demonstrating the utilization of our size-driven features for enhancing an existing real \ebay item recommendation service. 

\subsection{Experimental Setup}

\subsubsection{Dataset} \label{dataset_section}
To empirically validate our model, we collect a large-scale dataset sampled from fashion purchases done between 6/1/2019 and 6/1/2020 on the US domain of \ebay website. Our data consists of over 27M purchases made by 2.7M buyers and spans 210 fashion categories that include clothing, shoes and accessories categories. We next describe how we obtain and clean this data.

Since \ebay\ is not a dedicated fashion site, we retain only buyers with some minimum interest in fashion items -- those who have at least 5 fashion purchases over the given time period. For each item, we collect a set of attributes listed by the seller. These attributes include a free-text title (e.g., ``\emph{Men's Trainer Sneaker shoes, Sports Gym Casual Trainers, Outdoor Sneakers}''), a category structured as a leaf in the platform's category tree (e.g. `\texttt{men:men's shoes:sneakers}') and a list of attribute name-value pairs (e.g., `\texttt{department:men}', `\texttt{style:casual}', etc.).

Our platform allows sellers to list arbitrary attribute names and values. Hence, item attributes in our dataset are semi-structured with thick (and often noisy) long tails of attribute names and values. An implication of this is that, while most sellers (65\%) list a general `\texttt{size}' attribute (name), some sellers use a plethora of other attributes names such as `\texttt{men's size}', `\texttt{shoe size}', `\texttt{women shoe size}'; each applying to between 0.1\% and 5\% of our data. To complicate things further, there is no strict boundary between different attributes. In particular, the `\texttt{size}' attribute conflates with all other attribute names and often contains more data of specific size categories (e.g. `\texttt{shoe sizes}') than the more specific attributes (e.g. `\texttt{shoe size}'). We, therefore, keep the general `\texttt{size}' attribute along with 20 of the most common variations. However, to focus our experiments, we consider only the general `\texttt{size}' attribute in our evaluations. Along with the size attributes, we include 11 other potentially useful attributes such as \texttt{brand} and \texttt{gender}. Table \ref{table:context_attributes} lists all context attributes, and the percentage of samples they apply to.

\begin{table}[t]
\caption{Context attributes and their coverage (\% of samples they apply to)}
\small
\begin{tabular}{lc|lc}
\toprule
\textbf{Name} & \textbf{Coverage} & \textbf{Name} & \textbf{Coverage} \\
\midrule
Title      & 100\% &  Brand & 44.0\% \\
Category   & 100\% &  Occasion & 29.7\% \\
Department & 74.7\% & Manufacture Country  & 22.8\% \\
Brand Type & 60.0\% &   Fabric Type & 14.1\% \\
Style      & 59.6\% & Season & 10.7\% \\
Material   & 53.1\% & Gender & 6.8\% \\
Type       & 51.8\% &  &  \\
\toprule
\end{tabular}
\label{table:context_attributes}

\end{table}

A second outcome of our lax, semi-structured data, is having a long tail of attribute values. In particular the `\texttt{size}' attribute contains over 125K unique strings, most of which are either completely uninformative (e.g. `\texttt{one size}', `\texttt{not applicable}') or belong to a plethora of variation and spelling mistakes of common size measurements (e.g. `\texttt{xl}', `\texttt{extra large}', `\texttt{l-large}', `\texttt{32womens}', `\texttt{men 34}'). We filter out such noisy values so our evaluations focus on size personalization, rather than forcing our model to sort out naming variations of essentially the same size. To facilitate this, we first filter a closed list of uninformative strings which do not describe a specific size measurement (`\texttt{one size}', `\texttt{fits all}', `\texttt{not applicable}' etc.). We next merge abbreviated forms of textual sizes with their elongated forms, e.g., `\texttt{2xl}', `\texttt{xxl}' and `\texttt{extra extra large}', as these are very common variations. 
We then examine each category and for each attribute keep only unique values that appear at least 500 times and constitute at least 1\% of all the values within that category.
Finally, we retain only purchases where the purchased item has a size measurement listed by the seller. 
This aggressive filter results in removing 22\% of our size labels spread over 10's of thousands of unique strings, leaving 286 unique size labels. 

Another distinctive feature of our data is a high degree of data sparsity. We find that 92\% of the items are unique, constituting 58\% of all purchases, appear only once within the entire dataset. In addition, over 78\% of our buyers have less than 10 training purchases. These statistics suggest that during test time our model is likely to encounter many items it did not see before and many buyers which are only observed a few times in the training data.

\subsubsection{Baselines}
We evaluate \model against two strong baselines~\cite{AttentionRecsys2020,Sheikh2019a} which were previously reported to provide the best performance for this task. We further design three heuristic baselines to compare against \model. All heuristics consider the most granular category an item belongs to and back-off up the category tree to less granular categories if we do not have any data for the more granular category.

\noindent $\bullet$ \textbf{Most Common Value (MCV)}:
This is a simple baseline that does not do any personalization. MCV returns $P(s|c)$ -- the marginal probability of each size $s$ within a given category $c$. MCV effectively selects the most common size from the most granular category of the target item.

\noindent $\bullet$ \textbf{Most Recent Value (MRV)}:
This baseline considers the buyer's purchase history and selects the most recently purchased size within the target item's category. Adhering to our back-off strategy, we start by looking at purchases within the most granular category of the target item, and back-off to less granular categories if the buyer did not purchase any item from the more granular category.

\noindent $\bullet$ \textbf{Personalized Most Common Value (PMCV)}:
This baseline is a personalized version of MCV and considers the marginal probability of each size within the buyer's purchase history. Similar to the previous two baselines, PMCV looks only at items purchased from the same category as the target item and backs-off to less granular categories if no purchase is found.

\noindent $\bullet$ \textbf{SFNet}~\cite{Sheikh2019a}: This baseline implements a deep neural-network that learns item and buyer embeddings, which are further joined with item-based explicit features. To this end, both buyer and item features go through a series of non-linear layers with skip connections. The intermediate embeddings are then concatenated and go through another series of non-linear layers with skip connections. 

Unfortunately, the original SFNet implementation does not handle cold-start cases where a given item does not appear at all in the training data. Hence, to overcome this limitation, we restrict SFNet’s item embedding dictionary to include only item ids that appear least 2 times in our training data and map all other item ids into a `\texttt{[MISSING]}' embedding. Furthermore, we adjust some of the hyper-parameters suggested by the authors to ones that work better for our data. Specifically, we replace all tanh activations with ReLU activations, increase the embedding dimension to 64, increase the dimensions of the item and buyer pathways to (100, 50, 25) and the combined pathway to (50, 100, 200, 500). We further remove the L2 regularization, which we found not to be useful for our data.

\noindent $\bullet$ \textbf{Attention}~\cite{AttentionRecsys2020}: 
This baseline uses an encoder-decoder framework based on Transformers~\cite{vaswani2017attention}. A buyer's history is represented as a sequence of item ids, sizes and temporal embeddings that are fed into an encoder Transformer, and the outputs of this encoder are then concatenated with the target item id embedding and fed into a decoder Transformer. Similar to \model, this baseline uses a Transformer to attend to items in the buyer's history, yet with the main difference of not using any explicit content-based features. As in the case of SFNet, we restrict the item-id dictionary to items appearing at least twice in the training data.

\noindent $\bullet$ \textbf{Attention+}: 
During our experiments we found that the Attention baseline on its own does not perform well on our data. We hypothesize this is due to the difficulty of learning implicit features from our strongly sparse data. To test this hypothesis we replace item id embeddings used by the Attention baseline with category embeddings. This baseline can be viewed as a hybrid between \model and the Attention model. We denote this alternative baseline as Attention+ in our experiments. 

\subsubsection{Training and Hyper-parameters}
\label{sec:hyperparams}

\model requires a significant number of hyper-parameters to tune owing 
to the complex nature of our data and usage of Transformers. During our preliminary experiments with the validation set, we found that except the choice of embedding size and learning rate, other hyper-parameters have little effect on our results. We, therefore, leave all other hyper-parameters at their implementation defaults or coarsely tuned.

\squeezeupabit
\paragraph{Tokenization}
\label{sec:hyperparams_tokenization}

Throughout our experiments, we use the HuggingFace\footnote{https://huggingface.co/} implementation of the Byte-Pair Encoding (BPE) tokenizer to tokenize our strings. BPE is a state-of-the-art tokenizer extensively used in conjecture with neural-networks for its ability to avoid out-of-vocabulary tokens by breaking down unknown words to smaller n-grams. After tokenization, we either truncate or use a `\texttt{[PAD]}' token to pad all sequences to $45$ tokens, as we found this length suffice for $97\%$ of items and provides a good performance/efficiency trade-off. Unless otherwise stated, we use up to $25$ most recent purchases in a buyer’s history, as we found this to be more than enough (a detailed evaluation is provided in Section~\ref{ablations}). 
\squeezeupabit
\paragraph{Architecture}
\label{sec:hyperparams_architecture}

Unless otherwise stated, we use a hidden layer dimension of $d=512$ in all embedding and attention layers; and set the Transformer feed-forward dimension to $4 \times d$. We use GELU activation both in the Transformer hidden layers and in the classifier layers; except in the final layer, where we use softmax to produce a probability-like distribution over class predictions. Both the item and history Transformers use 4 stacked Transformer layers and 8 attention heads. The final classifier module has hidden dimensions of $(2 \times d,d, d/2, n)$ where $n=286$ is the number of output classes.
\squeezeupabit
\paragraph{Training}
\label{sec:hyperparams_training}

We use Pytorch\footnote{https://pytorch.org/} for all our experiments. We use the Adam optimizer~\cite{KingmaB14} for Stochastic Gradient Decent (SGD) optimization and mini-batches of 128 samples. 
All optimizer hyper-parameters are left at their Pytorch defaults; except the learning rate,  which we empirically set to an initial value of 10e-5. To control our learning rate during training, every 1K training iterations we measure our model’s performance on a 15K sample of the validation set. Whenever the loss does not decrease for 10 consecutive measurements, we decrease the learning rate by a factor of 2. We halt training when the learning rate drops below 10e-7.

\begin{table}[tb]
\caption{Comparison of \model performance to the baselines on the size prediction task. \model outperforms all baselines. All bold results are statistically significant.}
\small
\begin{tabular}{@{}lcccc@{}}
\toprule
\multirow{2}{*}{} & \textbf{Micro} & \textbf{Macro} & \textbf{Macro} & \textbf{Macro}\\
& \textbf{Precision} & \textbf{Precision} & \textbf{Recall} & \textbf{F1}\\
\midrule
\textbf{\model}          & \textbf{50.8\%}   & \textbf{51.8\%}    & \textbf{47.1\%}    & \textbf{47.7\%}         \\
\textbf{Attention}          & 38.0\%      & 33.3\%    & 19.9\%    & 23.2\%          \\
\textbf{Attention+}         & 47.4\%    & 32.2\%    & 24.1\%    & 25.4\%          \\
\textbf{SFNet}              & 40.3\%    & 47.9\%    & 43.4\%    & 44.0\%          \\
\textbf{PMCV}     & 40.3\%    & 24.4\%    & 19.7\%    & 20.4\%          \\
\textbf{MRV}      & 36.3\%    & 23.0\%    & 18.6\%      & 19.1\%          \\ 
\textbf{MCV}      & 21.1\%    & 4.4\%     & 5.2\%     & 3.7\%          \\
\bottomrule
\end{tabular}
\label{table:main_experiment}
\end{table}

\subsubsection{Evaluation Protocol and Metrics}
We split our dataset over time: 
we reserve the last 5 days in the data as a test set, the 5 days prior to that as a validation set and the rest are used as a training set. We train \model and the baselines on the training set and use the validation set to control the learning rate. 
We evaluate prediction performance using standard multi-class classification metrics. We consider micro and macro-averaged versions of precision, recall and F1 metrics. Micro averaged metrics average over all instances and are, therefore, less sensitive to smaller categories. On the other hand, macro averaged metrics are averaged over categories, giving more weight to smaller categories. Due to class symmetry in multi-class classification, the micro averaged precision, recall and F1 metrics are equal and hence we only report micro precision. To evaluate statistical significance, we use paired Student's t-test (p<0.05) for micro precision and paired bootstrap test (p<0.05) for the macro metrics~\cite{dror-etal-2018-hitchhikers}. We further apply Bonferroni correction in all cases.

\subsection{Size Prediction Results}
We next describe the main results of comparing \model to the baselines, analyzing the results on three key dimensions: \emph{Departments and Item Types}, \emph{Generalization to Unseen Items} and \emph{Account Types}. 

\subsubsection{\model vs. Baselines}
\label{experiment_baselines}
We compare the performance of \model to the baselines 
in Table \ref{table:main_experiment}. 
Overall, \model significantly outperforms all baselines by wide margins. Compared to the next best performing baseline, \model has gained at least +7\% better performance in all quality metrics.  In particular, we find both SFNet~\cite{Sheikh2019a} and the Attention~\cite{AttentionRecsys2020} baselines, which rely on learning implicit features of either buyers or items, do not perform well on our data. On the other hand, our modified Attention baseline (Attention+) shows that, by replacing the sparse item ids with category ids, this baseline can achieve a much better generalization. This suggests that sparsity is indeed a major consideration with our data and that using explicit features, such as category, can address this problem.

\subsubsection{Departments and Item Types}
Large \ecom platforms, such as \ebay, host extremely diverse catalogs of items and cater to diverse groups of buyers which do not necessarily follow similar patterns. To gain a deeper insight into the challenges of size personalization, we would like to break down and examine our data along multiple axes. We consider two such axes: item departments and item types. We classify items to four departments: \texttt{Men's}, \texttt{Women's}, \texttt{Unisex} and \texttt{Kid's}, where the last is anywhere from toddlers to youth. In addition, we classify items by five types: \texttt{Tops}, \texttt{Bottoms}, \texttt{Dress/Skirt}, \texttt{Footware} and \texttt{Other}.

\subsubsection{Generalization to Unseen Items} Given the aforementioned classification, a second question arises: can \model make size predictions for novel (unseen) items? Specifically, an interesting question is whether \model can generalize from one item type to another, and from one department to another. To evaluate this, we separate cases where a buyer bought an item from a 'Novel' department or item type which she did not purchase from before, forcing our model to generalize from one type of items to a different one (e.g., from `\texttt{shoe size}' to `\texttt{shirt size}', or from \texttt{Men's} to \texttt{Women's}). We contrast these from 'Observed' cases, where the buyer did purchase an item from the same department or item type.

\begin{table*}
\small
\caption{Evaluation by item categories, on observed and novel item types. Per department and item type, the underlined and bold values denote the best results for the observed and novel cases, respectively. * denotes statistically significant results.}
\setlength\tabcolsep{3pt}
\begin{tabular}{lcccccccccccc}
\toprule
 & \multicolumn{2}{c}{\textbf{PerSizE}} & \multicolumn{2}{c}{\textbf{Attention}} & \multicolumn{2}{c}{\textbf{Attention+}} & \multicolumn{2}{c}{\textbf{SFNet}} & \multicolumn{2}{c}{\textbf{PMCV}} & \multicolumn{2}{c}{\textbf{MCV}} \\
 & \multicolumn{1}{c}{\textbf{Observed}} & \multicolumn{1}{c}{\textbf{Novel}} & \multicolumn{1}{c}{\textbf{Observed}} & \multicolumn{1}{c}{\textbf{Novel}} & \multicolumn{1}{c}{\textbf{Observed}} & \multicolumn{1}{c}{\textbf{Novel}} & \multicolumn{1}{c}{\textbf{Observed}} & \multicolumn{1}{c}{\textbf{Novel}} & \multicolumn{1}{c}{\textbf{Observed}} & \multicolumn{1}{c}{\textbf{Novel}} & \multicolumn{1}{c}{\textbf{Observed}} & \multicolumn{1}{c}{\textbf{Novel}} \\
\midrule

\textbf{Mens}        & \underline{57\%*}   & 31.4\%*          & 43.5\% & 20.5\% & 53.5\% & 25.7\%          & 45.4\% & \textbf{31.9\%}* & 45.4\% & 24.1\% & 24.9\% & 23.9\% \\
\textbf{Womens}      & \underline{49.8\%*} & \textbf{29.7\%*} & 37.5\% & 17.5\% & 47.2\% & 26.4\%*          & 39.1\% & 27.7\%*          & 40.6\% & 22.0\%   & 20.1\% & 21.9\% \\
\textbf{Kids}        & \underline{41.9\%*} & \textbf{26.2\%*} & 24.1\% & 7.6\%  & 38.4\% & 19.8\%          & 29.3\% & 22.0\%*          & 29.7\% & 14.5\% & 9.0\%    & 13.9\% \\
\textbf{Unisex}      & \underline{56.4\%*} & \textbf{51.0\%*} & 42.0\% & 32.8\% & 45.8\% & 32.7\%          & 50.9\%* & 46.7\%*          & 41.0\%   & 25.3\% & 25.0\%   & 23.1\% \\
\midrule \midrule
\textbf{Tops}        & \underline{57.4\%*} & \textbf{35.7\%*} & 51.3\% & 22.6\% & 56.3\%* & 34\%*            & 46.7\% & 27.7\%          & 51.2\% & 17.6\% & 25.8\% & 25.3\% \\
\textbf{Bottoms}     & \underline{52.2\%*} & \textbf{29.2\%*} & 29.2\% & 15.4\% & 49.4\% & 22.2\%          & 35.5\% & 24.2\%          & 42.1\% & 14.1\% & 10.3\% & 11.5\% \\
\textbf{Dress/Skirt} & \underline{46.7\%*} & \textbf{32.3\%*} & 39.5\% & 23.9\% & 44.6\% & 28.2\%*          & 36.0\% & 24.3\%          & 39.6\% & 20.5\% & 17.5\% & 16.5\% \\
\textbf{Footware}    & \underline{49.4\%*} & 19.1\%*          & 16.7\% & 2.1\%  & 48.8\%* & \textbf{20.3\%}* & 29.2\% & 16.7\%*          & 37.3\% & 4.0\%  & 16.0\% & 14.7\%* \\
\bottomrule
\end{tabular}
\label{table:by_category}
\end{table*}

\begin{table*}
\small
\caption{Size prediction performance 
by account types. All bold results are statistically significant.}
\begin{tabular}{clcccccc}
\toprule
  &   & \textbf{\model} & \textbf{Attention} & \textbf{Attention+} & \textbf{SFNet} & \textbf{PMCV} & \textbf{MCV} \\
\midrule
\multirow{2}{*}{\textbf{\begin{tabular}[c]{@{}c@{}}Single Gender \\ 
Accounts\end{tabular}}}    & \textbf{Men's Items}   & \textbf{58.2\%}    & 45.1\%   & 53.9\%    & 46.8\%  & 45.7\%   & 24.6\%  \\
& \textbf{Women's Items}                            & \textbf{49.8\%}   & 37.3\%  & 46.6\%    & 38.7\%  & 40.0\%     & 19.5\%   \\
\midrule
\multirow{2}{*}{\textbf{\begin{tabular}[c]{@{}c@{}}Mixed Gender \\ 
Accounts\end{tabular}}}    & \textbf{Men's Items}   & \textbf{53.6\%}    & 39.1\%   & 50.5\%    & 42.4\%  & 42.9\%   & 25.0\%   \\
& \textbf{Women's Items}                            & \textbf{48.0\%}    & 35.1\%   & 45.7\%    & 38.3\%  & 39.0\%     & 19.6\%   \\
\midrule
\midrule
\multirow{2}{*}{\textbf{\begin{tabular}[c]{@{}c@{}}Single Age Group \\ 
Accounts\end{tabular}}} & \textbf{Adult Items}      & \textbf{52.4\%}    & 39.9\%   & 49.0\%    & 41.9\%  & 42.1\%   & 22.0\%  \\
& \textbf{Kids Items}                               & \textbf{43.6\%}    & 31.7\%  & 38.5\%    & 31.3\%  & 31.0\%   & 7.9\%    \\
\midrule
\multirow{2}{*}{\textbf{\begin{tabular}[c]{@{}c@{}}Mixed Age Group \\ 
Accounts\end{tabular}}}  & \textbf{Adult Items}     & \textbf{49.0\%}    & 34.6\%   & 45.9\%    & 38.1\%  & 38.6\%   & 22.4\%  \\
& \textbf{Kids Items}                               & \textbf{41.4\%}    & 22.2\%   & 38.2\%    & 28.9\%  & 29.2\%   & 9.4\%   \\
\bottomrule    
\end{tabular}
\captionsetup{width=0.8\textwidth}
\label{table:by_account_type}
\end{table*}

We summarize \model and baselines performance over these item groups in Table~\ref{table:by_category}. Looking at the 'Observed' column, we can see that while \model consistently outperforms all baselines, not all departments and item types behave the same. Specifically, \texttt{Kid's} items are significantly more challenging than all adult items, while \texttt{Men's} sizes are the least challenging. We further observe a significant variance among item types, where \texttt{Tops} are the least challenging and \texttt{Footware} are the most. Such differences can be attributed to the nature of the data itself, where some categories (e.g., \texttt{Shoes}) are more nuanced than others (e.g., \texttt{Shirts}). 

Looking at the 'Novel' column, we see that, the performance of \model and all baselines significantly drops  in these challenging situations. However, \model significantly outperforms the baselines in almost all cases.
In particular, \model outperforms the non-personalized MCV heuristic, suggesting that the former can indeed generalize to unseen purchased categories. A notable observation is further made for \texttt{Footware} items, where \model struggles to generalized from other clothing items into footware items, but still manages to perform some personalization and beats the non-personalized MCV heuristic by 4.4\%. Another interesting observation is made for the Attention baseline, which performs poorly on Novel cases with the lack of explicit features. On the other hand, those baselines that do use explicit features (i.e., Attention+, SFNet) perform much better. These results demonstrate the importance of using explicit features to improve generalization in such settings.

\subsubsection{Account Types}
Next, we examine our data from the target audience demographic perspective and divide it by apparent gender and age group of buyer accounts. We define apparent gender and age group of an account in terms of the item categories in its purchase history. `\texttt{Kids}' accounts are, therefore, accounts purchasing only from the kids category; while `\texttt{Adult}' men and women accounts are accounts purchasing only from men's and women's categories, respectively. Accounts purchasing from both men's and women's categories are termed `\texttt{Mix-Gender}'; while accounts purchasing from both kids and adults categories are termed `\texttt{Mixed-Age Group}'. We summarize \model and baselines performance by account and item categorizations in Table \ref{table:by_account_type}. Overall, \model performs well on all account categorizations. As we would expect, \model performance degrades for mixed accounts, yet it still performs well in all cases and outperforms all baselines. We attribute \model's success to its ability to attend to specific purchases within the buyer's history based on features such as category and gender.

\subsection{Ablation Study} \label{ablations}
We next perform an extensive set of experiments and ablations of \model to identify which components and features contribute the most to it's performance. Throughout this section, all results are obtained by training each model five times using different seeds to randomly initialize model weights and reporting the average performance. We do this in order to reduce the variance of reported numbers, as we found that our results may vary by as much as 1\% for different seeds.

\subsubsection{Embedding Size}
We vary the embedding dimension of the model, which effectively determines its size in terms of parameter count. We report results in Table~\ref{table:embed}, where we can clearly see that, increasing the embedding dimension significantly improves \model performance. In particular, we find large improvements in Macro F1, as the embedding dimension increases. This shows that larger models are particularly better at fitting less frequent categories.

\subsubsection{Purchase History Length}
\begin{figure}[tbh]
  \centering
  \includegraphics[scale=0.6]{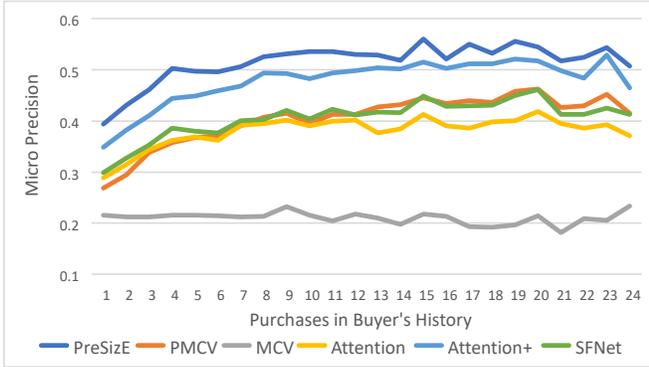}
  \caption{\model performance by history length}
   \label{fig:history}
\end{figure}

We plot in Figure~\ref{fig:history} the impact of the number of items in a buyer's history on model performance. We can see that, \model can identify the correct size with 40\% precision using as little as a single item in the buyer's purchase history. However, precision improves significantly with more history items and peaks at around 15 items. Comparing \model to baselines, and especially Attention+, shows that \model performs better both overall and particularly when the buyer has less items in her purchase history.       

\subsubsection{Component Analysis}
We continue to perform an ablation by removing different parts of \model and observing the impact on its performance. 
Due to the large number of models fitted in this experiment, and to reduce cost and resource usage, we use a lowered embedding dimension size of 128 throughout the experiment. 

\begin{table}[tbh]
\small
\captionsetup{width=0.45\textwidth}
\caption{\model performance by embedding dimension. Model parameter counts (model size) are noted as well.}
\begin{tabular}{cccc}
\toprule
\multicolumn{1}{c}{\textbf{\begin{tabular}[c]{@{}c@{}}Embedding \\ size\end{tabular}}} & \multicolumn{1}{c}{\textbf{\begin{tabular}[c]{@{}c@{}}Parameter \\ Count\end{tabular}}} &
\multicolumn{1}{c}{\textbf{\begin{tabular}[c]{@{}c@{}}Micro \\ Precision\end{tabular}}} & \multicolumn{1}{c}{\textbf{\begin{tabular}[c]{@{}c@{}}Macro \\ F1\end{tabular}}} \\
\midrule
\textbf{32}   & 1.1M & 44.5\%  & 20.3\%   \\
\textbf{64}   & 2.5M & 47.8\%  & 34.7\%   \\
\textbf{128}  & 6M  & 49.2\%  & 41.0\%   \\
\textbf{256}  & 16M  & 50.3\%  & 45.3\%   \\
\textbf{512}  & 48M  & \textbf{50.8\%}  & \textbf{47.7\%}  \\
\bottomrule
\end{tabular}
\label{table:embed}
\end{table}

We report the results of the ablation in Table~\ref{table:ablation}. 
First, we observe that, rich context features are essential for good performance. Moreover, removing the temporal embeddings degrades the performance by 0.5\%. This suggests that, \model can (at least partially) model temporal changes in size preferences. 

Among the rest of our features, 
\texttt{Category}, \texttt{Brand} and \texttt{Style} are the most impactful. In particular, using \texttt{Category} alone achieves 46.2\% and the rest of the features amounts to 3\% additional improvement. Yet, removing any single feature does not degrade performance by much. This suggests that, no single feature contains information that cannot be mostly extracted from other features  (e.g., \texttt{Department} and \texttt{Gender} can be deduced from the \texttt{Category}). 

\subsection{Using Size Prediction for Similar Items Recommendation}\label{recommend with size}
We conclude this section by demonstrating the merits of utilizing \model's derived size prediction features (see again Section~\ref{sec:features}) for enhancing a real downstream \ebay recommendation service. To this end, we perform an offline evaluation of \ebay's similar items recommendation service, focusing on the Fashion domain.

\begin{table}[tb]
\small
\caption{Ablation study of \model, where we remove each part/feature of the model and observe effects on its performance. Numbers in parenthesis denote the opposite: \model performance when removing all features but one.}
\begin{tabular}{lc}
\toprule
\textbf{Part of Model/Feature removed}  & \textbf{Micro-Precision} \\
\midrule
\textbf{All Features (No Features)} & 49.2\% (36.4\%) \\
\textbf{-- All Context Features} & 36.5\% (-) \\
\textbf{-- Temporal Embedding} & 48.7\% (-) \\
\midrule
\textbf{-- Category} & 48.3\% (46.2\%) \\
\textbf{-- Brand} & 48.8\% (39.2\%) \\
\textbf{-- Style} & 48.9\% (42.3\%) \\
\textbf{-- Occasion} & 49.1\% (36.7\%) \\
\textbf{-- Country of Manufacture} & 49.1\% (36.3\%) \\
\textbf{-- Fabric Type} & 49.1\% (37.4\%) \\
\textbf{-- Department} & 49.2\% (37.7\%) \\
\textbf{-- Gender} & 49.2\% (36.5\%) \\
\textbf{-- Material} & 49.2\% (37.5\%) \\
\textbf{-- Type} & 49.2\% (40.9\%) \\
\textbf{-- Title} & 49.2\% (36.4\%) \\
\textbf{-- Brand-Type} & 49.3\% (36.7\%) \\
\textbf{-- Season} & 49.3\% (36.2\%) \\
\bottomrule
\end{tabular}
\label{table:ablation}
\end{table}

Most \ecom websites have an item page containing a module of \emph{Similar Items} which recommends items that are similar to the featured item.
The existing model for similar items recommendation on the item page at \ebay considers hundreds of buyer-based, seller-based and item-based features. We next propose to extend the buyer-based features with the additional set of size prediction features (\emph{total score},\emph{best score} and \emph{best rank}). The recommender's model is optimized for item purchases, ranking higher items that are more likely to be purchased by the buyer. 

We next compare the performance of the similar items recommendation model when adding the size prediction features against the existing model. 
To further assess the advantage of using a complex model such as \model, we compare \model-based features to similar features extracted using an effective, yet simpler, heuristic. Specifically we extract the same features from the PMCV baseline, which was shown in Section \ref{experiment_baselines} to be an effective baseline, and compare the same item recommendation model trained with these features.
We sample from two weeks of item impressions data between 10/17/2020 and 10/30/2020 from the logs of \ebay's similar items recommendation module, one for training and one for held-out evaluation. The size prediction features are calculated for each buyer in our data based on her history of purchases in the last year prior to the two-week experiment period. We consider several recommendation quality metrics. The first is the relative gain in \emph{Sale Rank}: the average rank of the top purchased item in the results set. The two other metrics are the relative gain in \emph{Purchase-Through Rate} (PTR) and \emph{Normalized Discounted Cumulative Gain} (NDCG), both measured considering the top-5 results.

We report the results of our evaluation in Table~\ref{table:plsim}. As we can observe, using size prediction features derived from \model provides significantly better recommendation quality then using those derived from PMCV.  

\begin{table}[tbh]
\caption{Gain in item recommendation quality with the PMCV baseline size features and \model features compared to not using any size features}
\small
\begin{tabular}{rrrr}
\toprule
\multicolumn{1}{c}{\textbf{\begin{tabular}[c]{@{}c@{}}Method\end{tabular}}} & \multicolumn{1}{c}{\textbf{\begin{tabular}[c]{@{}c@{}}Sale Rank\end{tabular}}} & \multicolumn{1}{c}{\textbf{\begin{tabular}[c]{@{}c@{}}PTR@5 \end{tabular}}} & \multicolumn{1}{c}{\textbf{\begin{tabular}[c]{@{}c@{}}NDCG@5 \end{tabular}}} \\
\midrule
\textbf{PMCV Features}                                                                           & +0.73\%                                                                                 & +0.49\%  & +0.41\%                                                                           \\
\textbf{\model Features}                                                                           & +2.10\%                                                                                 & +2.02\%    & +1.57\%                                                                        \\
\bottomrule
\end{tabular}
\label{table:plsim}
\end{table}

\section{Conclusions}
We proposed \model, a deep learning framework designed to predict buyer's size preference in \ecom. \model uses Transformers for content-based and sequence-based size prediction. To validate \model, we performed a large array of experiments on \ecom data, showing that its performance transcends that of state-of-the-art baselines. 
We attribute \model's success to effective usage of explicit content-based features, and support this by an extensive ablation experiment.

We further demonstrated the importance of size prediction in improving downstream item recommendation services in \ecom. 
Specifically, we used \model's size predictions as features to improve an existing learning-to-rank item recommendation service at eBay. We showed that, these features can significantly improve the existing service's performance as well as outperform size features generated by a strong heuristic baseline. 
\balance

\bibliographystyle{ACM-Reference-Format}
\bibliography{sizes}

\end{document}